\documentclass[twocolumn,aps,showpacs,superscriptaddress]{revtex4}
\usepackage{graphicx}
\usepackage{dcolumn}
\usepackage{bm}

\begin{document}

\def\Journal#1#2#3#4{{#1} {\bf #2}, #3 (#4)}

\def\NCA{Nuovo Cimento}
\def\NIM{Nucl. Instr. Meth.}
\def\NIMA{{Nucl. Instr. Meth.} A}
\def\NPB{{Nucl. Phys.} B}
\def\NPA{{Nucl. Phys.} A}
\def\PLB{{Phys. Lett.}  B}
\def\PRL{Phys. Rev. Lett.}
\def\PRC{{Phys. Rev.} C}
\def\PRD{{Phys. Rev.} D}
\def\ZPC{{Z. Phys.} C}
\def\JPG{{J. Phys.} G}
\def\CPC{Comput. Phys. Commun.}
\def\EPJ{{Eur. Phys. J.} C}

\preprint{}

\title{Identified hadron compositions in p+p and Au+Au collisions at high transverse momenta at $\sqrt{s_{_{NN}}} = 200$ GeV}

\affiliation{Argonne National Laboratory, Argonne, Illinois 60439,
USA} \affiliation{University of Birmingham, Birmingham, United
Kingdom} \affiliation{Brookhaven National Laboratory, Upton, New
York 11973, USA} \affiliation{University of California, Berkeley,
California 94720, USA} \affiliation{University of California,
Davis, California 95616, USA} \affiliation{University of
California, Los Angeles, California 90095, USA}
\affiliation{Universidade Estadual de Campinas, Sao Paulo, Brazil}
\affiliation{University of Illinois at Chicago, Chicago, Illinois
60607, USA} \affiliation{Creighton University, Omaha, Nebraska
68178, USA} \affiliation{Czech Technical University in Prague,
FNSPE, Prague, 115 19, Czech Republic} \affiliation{Nuclear
Physics Institute AS CR, 250 68 \v{R}e\v{z}/Prague, Czech
Republic} \affiliation{University of Frankfurt, Frankfurt,
Germany} \affiliation{Institute of Physics, Bhubaneswar 751005,
India} \affiliation{Indian Institute of Technology, Mumbai, India}
\affiliation{Indiana University, Bloomington, Indiana 47408, USA}
\affiliation{Alikhanov Institute for Theoretical and Experimental
Physics, Moscow, Russia} \affiliation{University of Jammu, Jammu
180001, India} \affiliation{Joint Institute for Nuclear Research,
Dubna, 141 980, Russia} \affiliation{Kent State University, Kent,
Ohio 44242, USA} \affiliation{University of Kentucky, Lexington,
Kentucky, 40506-0055, USA} \affiliation{Institute of Modern
Physics, Lanzhou, China} \affiliation{Lawrence Berkeley National
Laboratory, Berkeley, California 94720, USA}
\affiliation{Massachusetts Institute of Technology, Cambridge, MA
02139-4307, USA} \affiliation{Max-Planck-Institut f\"ur Physik,
Munich, Germany} \affiliation{Michigan State University, East
Lansing, Michigan 48824, USA} \affiliation{Moscow Engineering
Physics Institute, Moscow Russia} \affiliation{NIKHEF and Utrecht
University, Amsterdam, The Netherlands} \affiliation{Ohio State
University, Columbus, Ohio 43210, USA} \affiliation{Old Dominion
University, Norfolk, VA, 23529, USA} \affiliation{Panjab
University, Chandigarh 160014, India} \affiliation{Pennsylvania
State University, University Park, Pennsylvania 16802, USA}
\affiliation{Institute of High Energy Physics, Protvino, Russia}
\affiliation{Purdue University, West Lafayette, Indiana 47907,
USA} \affiliation{Pusan National University, Pusan, Republic of
Korea} \affiliation{University of Rajasthan, Jaipur 302004, India}
\affiliation{Rice University, Houston, Texas 77251, USA}
\affiliation{Universidade de Sao Paulo, Sao Paulo, Brazil}
\affiliation{University of Science \& Technology of China, Hefei
230026, China} \affiliation{Shandong University, Jinan, Shandong
250100, China} \affiliation{Shanghai Institute of Applied Physics,
Shanghai 201800, China} \affiliation{SUBATECH, Nantes, France}
\affiliation{Texas A\&M University, College Station, Texas 77843,
USA} \affiliation{University of Texas, Austin, Texas 78712, USA}
\affiliation{University of Houston, Houston, TX, 77204, USA}
\affiliation{Tsinghua University, Beijing 100084, China}
\affiliation{United States Naval Academy, Annapolis, MD 21402,
USA} \affiliation{Valparaiso University, Valparaiso, Indiana
46383, USA} \affiliation{Variable Energy Cyclotron Centre, Kolkata
700064, India} \affiliation{Warsaw University of Technology,
Warsaw, Poland} \affiliation{University of Washington, Seattle,
Washington 98195, USA} \affiliation{Wayne State University,
Detroit, Michigan 48201, USA} \affiliation{Institute of Particle
Physics, CCNU (HZNU), Wuhan 430079, China} \affiliation{Yale
University, New Haven, Connecticut 06520, USA}
\affiliation{University of Zagreb, Zagreb, HR-10002, Croatia}

\author{G.~Agakishiev}\affiliation{Joint Institute for Nuclear Research, Dubna, 141 980, Russia}
\author{M.~M.~Aggarwal}\affiliation{Panjab University, Chandigarh 160014, India}
\author{Z.~Ahammed}\affiliation{Variable Energy Cyclotron Centre, Kolkata 700064, India}
\author{A.~V.~Alakhverdyants}\affiliation{Joint Institute for Nuclear Research, Dubna, 141 980, Russia}
\author{I.~Alekseev}\affiliation{Alikhanov Institute for Theoretical and Experimental Physics, Moscow, Russia}
\author{J.~Alford}\affiliation{Kent State University, Kent, Ohio 44242, USA}
\author{B.~D.~Anderson}\affiliation{Kent State University, Kent, Ohio 44242, USA}
\author{C.~D.~Anson}\affiliation{Ohio State University, Columbus, Ohio 43210, USA}
\author{D.~Arkhipkin}\affiliation{Brookhaven National Laboratory, Upton, New York 11973, USA}
\author{G.~S.~Averichev}\affiliation{Joint Institute for Nuclear Research, Dubna, 141 980, Russia}
% ...

\collaboration{STAR Collaboration}\noaffiliation

\date{\today}

\begin{abstract}
We report transverse momentum ($p_{T}
 \leq15$ GeV/$c$) spectra of
$\pi^{\pm}$, $K^{\pm}$, $p$, $\bar{p}$, $K_{S}^{0}$, and
$\rho^{0}$ at mid-rapidity in p+p and Au+Au collisions at
$\sqrt{s_{_{NN}}}$ = 200 GeV. Perturbative QCD calculations are
consistent with $\pi^{\pm}$ spectra in p+p collisions but do not
reproduce $K$ and $p(\bar{p})$ spectra. The observed decreasing
antiparticle-to-particle ratios with increasing $p_T$ provide
experimental evidence for varying quark and gluon jet
contributions to high-$p_T$ hadron yields. The relative hadron
abundances in Au+Au at $p_{T}{
 }^{>}_{\sim }8$ GeV/$c$ are measured to be similar to the
p+p results, despite the expected Casimir effect for parton energy
loss.

\end{abstract}

 \maketitle

Quarks and gluons are the fundamental particles carrying color
charge and participating in the strong interaction. High-energy
partons are produced through hard processes in hadron-hadron
collisions and, like all particles carrying color or electric
charges, lose energy while traversing the hot and dense medium
created in heavy ion collisions~\cite{jetquench,whitepapers}. In
all model calculations, the amount of parton energy loss is
proportional to the color-charge Casimir factor (the relative
coupling strength of gluon radiation from quarks or from gluons),
and strongly depends on the medium traversed and on the parton
mass~\cite{jetquench,renk:07,xinnian:98}. This energy loss
suppresses hadron spectra at high $p_T$ in heavy ion collisions,
an effect referred to as jet quenching and quantified by the
nuclear modification factors ($R_{AA}$, the ratio of heavy ion
collision spectra to p+p collision spectra scaled by the number of
underlying binary nucleon-nucleon inelastic collisions)
~\cite{jetquench,starhighpt,rhicotherhighpt}.

The study of identified hadron spectra at high $p_T$ in p+p
collisions also provides quantitative constraints on model
calculations based on perturbative quantum chromodynamics
(pQCD)~\cite{pQCD}. In next-to-leading order (NLO) pQCD
calculations, inclusive production of single hadrons is described
by the convolution of parton distribution functions (PDFs),
parton-parton interaction cross sections, and fragmentation
functions (FFs). Specifically, the FFs~\cite{AKK,AKK2008,DSS} were
primarily derived from elementary electron-positron collisions.
The NLO pQCD framework has been verified with calculations
successfully describing the spectra of inclusive charged hadrons,
$\pi^{0}$, and jets~\cite{starhighpt,rhicotherhighpt,starjet} at
RHIC. However, the flavor separated quark and gluon FFs are not
well constrained, especially for baryon production. To understand
further the mechanisms of particle production in p+p collisions
and parton interactions with the medium in heavy ion collisions,
it is necessary to provide more stringent constraints on the quark
and gluon FFs by comparing theoretical calculations with
experimental data in the same kinematics in p+p collisions.

Measurements sensitive to the flavor of the initial hard scattered
parton will provide further constraints and insights into the jet
quenching
mechanism~\cite{jetquench,renk:07,xinnian:98,urs:07,Fries:08,starAuAuPID,weiliu:07,enko:10,color:11}.
An open question is whether the interaction of the hard partons
with the medium alters the relative abundances of the identified
particle spectra (jet chemistry). Two examples of these
interactions with the medium are enhanced parton
splitting~\cite{urs:07} and flavor changes of the initial parton
(jet conversion)~\cite{Fries:08}. These processes are expected to
modify the high-$p_T$ identified particle ratios in heavy ion
versus p+p collisions. The centrality dependence of antiproton and
pion spectra in Au+Au collisions indicates that the suppression
magnitude for antiprotons is similar to that for
pions~\cite{starAuAuPID}. This is unexpected since antiproton
production is dominated by gluon fragmentation, while pions have a
comparable contribution from both gluon and quark jets~\cite{AKK}.
The Casimir factor for gluons is 9/4 times that for quarks, which
is expected to induce larger energy loss when gluons traverse the
medium~\cite{xinnian:98}. Naively, this would result in more
suppressed antiproton spectra compared to pion spectra. A jet
conversion mechanism, where a parton can change flavor or color
charge after interaction with a medium, has been proposed whose
calculations show a net quark to gluon jet conversion in this
medium~\cite{weiliu:07}. This leads to a better agreement with
experimental data~\cite{weiliu:07,enko:10}. It is also predicted
that the suppression pattern of kaons would differ significantly
from that of pions due to the notable difference in relative
abundance of strange quarks produced in jets versus the
statistical expectations in a hot and dense
medium~\cite{Fries:08}. Experimental measurements of identified
hadrons at high $p_T$ in p+p collisions are required to more
accurately determine the p+p reference and to provide further
constraints to the FFs. Together with the Au+Au measurements, it
will help to understand the parton interactions with the medium.

In this Letter, we report $\pi^{\pm}$, $K^{\pm}$, $p(\bar{p})$,
$K_S^{0}$, and $\rho^{0}$ $p_T$ spectra at mid-rapidity
($|y|\!\!<$ 0.5) up to 15 GeV/$c$ in p+p collisions at
$\sqrt{s_{_{NN}}}$ = 200 GeV. The hadron spectra and particle
ratios in p+p collisions are compared to NLO pQCD calculations
with various FFs. In addition, spectra of $K^{\pm}+p(\bar{p})$
(measured by $h^{\pm}-\pi^{\pm}$), $K_S^{0}$, and $\rho^{0}$ in
the 12\% most central Au+Au collisions are presented. $R_{AA}$ are
presented for $K^{\pm}+p(\bar{p})$, $K_S^{0}$, $\pi^{+}+\pi^{-}$,
and $\rho^{0}$.

A total of 21 million 12\% most central Au+Au collisions used in
this analysis were taken in 2004 at STAR~\cite{star}. The central
trigger was based on an on-line cut of energy deposited in the
Zero-Degree Calorimeters. The p+p data used for this analysis were
taken in 2005. Experimental study of identified hadrons at high
$p_T$ in p+p collisions was made possible by two technical
advances: (1) using the STAR Barrel Electro-Magnetic Calorimeter
(BEMC)~\cite{BEMC} as a trigger device for charged hadrons in p+p
collisions; and (2) improving the calibration and understanding of
the ionization energy loss ($dE/dx$) of charged particles in the
relativistic rise region in the Time Projection Chamber
(TPC)~\cite{startpc}. The minimum bias p+p collision events were
identified by the coincidence of two beam-beam
counters~\cite{starhighpt}. Online triggers, which utilized a
minimum bias trigger and the energy deposited in either a single
BEMC tower (high tower trigger, HT) or in a contiguous
$\Delta\eta\times\Delta\phi$ = $1\times1$ rad region (jet patch
trigger, JP) of the BEMC, were used for the p+p collisions. A
total of 5.6 million JP events with transverse energy $E_{T} > $
6.4 GeV were used for $\pi^{\pm}$, $K^{\pm}$, and $p(\bar{p})$
analyses. To reduce trigger biases, only away-side particles (at
azimuthal angles $90^{\circ}\!-\!270^{\circ}$ from the JP trigger)
were used in the analysis. Another 5.1 million events with $E_{T}
>$ 2.5 GeV (HT1), and 3.4 million events with $E_{T}>$ 3.6 GeV
(HT2) were used for $K_S^{0}\rightarrow\pi^{+}\pi^{-}$ and
$\rho^{0}\rightarrow\pi^{+}\pi^{-}$ reconstruction by requiring
that one of the daughter pions trigger the high tower. The trigger
enhancement factor in the range of
10--1000~\cite{reCalibrationMethod} and bias have been determined
by embedding PYTHIA events in the STAR geometry and selecting
events that pass various detector thresholds present in real
events. Consistencies of spectra from minimum bias datasets and
between charged and neutral hadrons in the overlapping $p_T$ range
were utilized to check the trigger corrections.

The $dE/dx$ measured in the TPC was used to identify $\pi^{\pm}$,
$K^{\pm}$, and $p(\bar{p})$ at $3\!<p_T\!<\!15$ GeV/$c$ at
mid-rapidity~\cite{pidNIMA,starAuAuPID,starppdAuPID}. The pion,
kaon, and proton yields were extracted from a three-Gaussian fit
to the inclusive positively or negatively charged particle $dE/dx$
distributions at a given momentum. The re-calibrated $dE/dx$ in
the TPC ~\cite{reCalibrationMethod} enabled us to measure
high-$p_T$ kaons. $K_S^{0}\rightarrow\pi^{+}+\pi^{-}$ decays were
identified through the V0 topology~\cite{starks}. The
$\rho^{0}\rightarrow\pi^{+}+\pi^{-}$ yields were obtained using
cocktail methods, after like-sign $\pi^{+}\pi^{+}$ and
$\pi^{-}\pi^{-}$ pair invariant mass distribution backgrounds were
subtracted from unlike-sign $\pi^{+}\pi^{-}$ pair
distributions~\cite{starrho}. For the line shape of
$\rho^{0}\rightarrow\pi^{+}+\pi^{-}$, the procedure and formula
in~\cite{starrho} were used with the $\rho^{0}$ mass at 775 MeV
and Breit-Wigner width 155 MeV~\cite{fachini:08}. The possible
$\sigma^{0}$ particle~\cite{sigma0:11} (mass at $\approx600$ MeV
and Breit-Wigner width scanning from 100 to 500 MeV) was included
in the cocktail fit as part of the systematic study on effect of
other contributions on $\rho^{0}$ yields. This results in
$\pm20$\% systematic error in $\rho^{0}$ yields and improves the
$\chi^{2}$ per degree of freedom ($\chi^{2}/NDF$) up to a factor
of 3 to be around unity. The fit with best $\chi^{2}/NDF$ was used
to obtain the default $\rho^{0}$ yields, where the
$\sigma^{0}/\rho^{0}$ ratio is about 25\% independent of $p_T$. An
additional systematic check was performed using the modified
Soeding parametrization for a possible interference
effect~\cite{UPCrhopaper} on $\rho^{0}$ line shape. This results
in larger $\chi^{2}/NDF$ and $\rho^{0}$ yields are within the
stated systematic uncertainty.

Acceptance and efficiency corrections were studied by Monte Carlo
GEANT simulations. Weak-decay feed-down contributions (e.g.
$K_S^{0}\rightarrow\pi^{+}+\pi^{-}$) are subtracted from the pion
spectra~\cite{starAuAuPID}. Inclusive $p$ and $\bar{p}$ production
are presented, without hyperon feed-down
subtraction~\cite{starAuAuPID}. In central Au+Au collisions,
systematic errors for $K_{S}^{0}$ yields are
4--10\%~\cite{starcucustrangeness}, and those for $\rho^{0}$
yields are 32\%, dominated by signal reconstructions (20\%) and
cocktail fits (20\%). The systematic errors from low to high $p_T$
for $\pi^{\pm}$, $K^{\pm}$, $p$, and $\bar{p}$ in p+p collisions
include uncertainties in efficiency ($\approx5$\%), $dE/dx$
position and width (5--70\%), momentum distortion due to charge
build-up in the TPC volume (0--12\%), the smearing of the measured
spectra due to momentum resolution (0--7\%), and trigger
correction factors (40--10\%). Systematic uncertainties for
$K^{0}_{S}$ and $\rho^{0}$ yields in p+p collisions include
uncertainties in trigger enhancement factors and biases
($\!<\!20$\%), momentum resolution (1--20\%), efficiency (5\%),
and cocktail fits of $\rho^{0}$ yields (20\%). The normalization
uncertainties on the invariant yields and cross sections are 8\%
and 14\% in p+p collisions, respectively. The cancellation of the
correlated systematic errors is taken into account for the
particle ratios.

\begin{figure}
\includegraphics*[keepaspectratio,scale=1.]{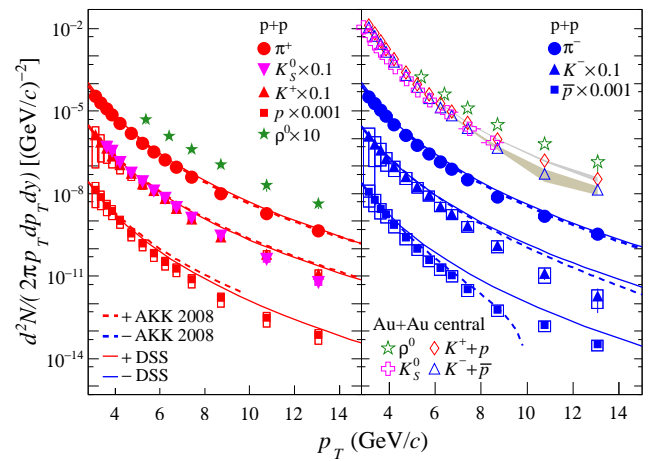}
\caption{(Color Online) The invariant yields
$d^2N/(2{\pi}p_Tdp_Tdy)$ of $\pi^{\pm}$, $K^{\pm}$, $K_{S}^{0}$,
$\rho^{0}$, $p$, and $\bar{p}$ from non-singly diffractive p+p
collisions ($\sigma_{_{NSD}}=30.0\pm3.5$ mb~\cite{starhighpt}),
those of $K+p(\bar{p})$, $K_{S}^{0}$, and $\rho^{0}$ in central
Au+Au collisions, and NLO calculations with AKK~\cite{AKK2008} and
DSS~\cite{DSS} FFs. The uncertainty of yields due to the scale
dependence as evaluated in~\cite{DSS} is about a factor of 2. Bars
and boxes (bands) represent statistical and systematic
uncertainties, respectively.} \label{spectra}
\end{figure}

The invariant yields $d^2N/(2{\pi}p_Tdp_Tdy)$ of $\pi^{\pm}$,
$K^{\pm}$, $K_{S}^{0}$, $\rho^{0}$, $p$, and $\bar{p}$ from p+p
collisions, and those of $K+p(\bar{p})$, $K_{S}^{0}$, and
$\rho^{0}$ in central Au+Au collisions are shown in
Fig.~\ref{spectra}. In p+p collisions, our measurements are
consistent with those from minimum bias collisions within
systematic errors in the overlapping $p_T$
region~\cite{starppdAuPID}. The $K^{\pm}$ and $K^{0}_{S}$ yields
are consistent within statistical and systematic uncertainties,
which verifies that the JP trigger condition for the $K^{\pm}$
measurement was correctly accounted for in the simulation. Also
shown in Fig.~\ref{spectra} are the NLO calculations for
$\pi^{\pm}$, $K^{\pm}$, $p$, and $\bar{p}$ spectra based on
AKK~\cite{AKK2008} and DSS~\cite{DSS} FFs. Both calculations are
consistent with the charged pion spectra in p+p collisions, but
deviate from the kaon and proton spectra.

\begin{figure}
\includegraphics*[keepaspectratio,scale=1.]{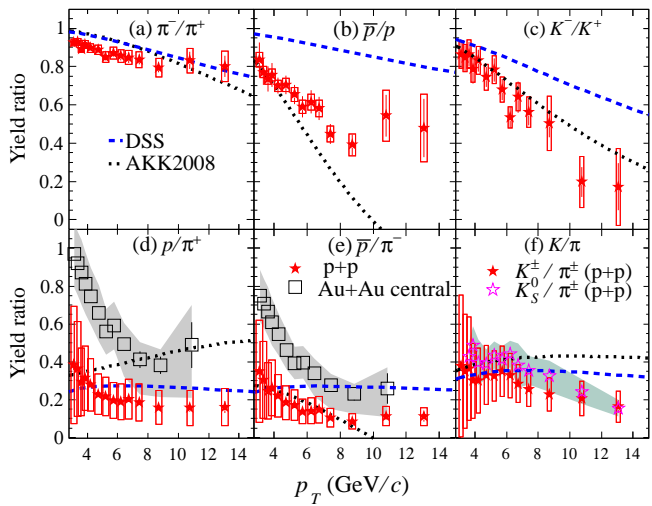}
\caption{(Color Online) Yield ratios $\pi^{-}$/$\pi^{+}$,
$\bar{p}/p$, $K^{-}$/$K^{+}$, $p/\pi^{+}$, $\bar{p}$/$\pi^{-}$,
and $(K^{\pm},K_{S}^{0})/\pi^{\pm}$ versus $p_T$ in p+p
collisions, and nominal NLO calculations with AKK~\cite{AKK2008}
and DSS~\cite{DSS} FFs without theoretical uncertainties. The open
squares in panels (d) and (e) are the $p/\pi^{+}$ and
$\bar{p}$/$\pi^{-}$ ratios in central Au+Au
collisions~\cite{starAuAuPID} with updated uncertainties at high
$p_T$, and all other data points are from p+p collisions. Bars and
boxes (bands) represent statistical and systematic uncertainties,
respectively.} \label{ratios}
\end{figure}

In Fig.~\ref{ratios}, particle ratios are shown as star symbols as
a function of $p_{T}$ from p+p collisions. Our results are
consistent with minimum bias results~\cite{starppdAuPID} in the
overlapping $p_T$ region and are extended to $p_T\approx15$
GeV/$c$. We show for the first time that at this collision energy,
$\pi^{-}$/$\pi^{+}$, $\bar{p}/p$, and $K^{-}$/$K^{+}$ ratios
decrease with increasing $p_T$ in p+p collisions at mid-rapidity.
This indicates relatively larger valence quark contributions to
$\pi^{+}$, $K^{+}$, and $p$ at high $p_T$ than to their respective
antiparticles. The NLO pQCD calculations with DSS and AKK FFs are
consistent with the $\pi^{-}$/$\pi^{+}$ ratio but deviate from
most of the other ratios measured. In the past, flavor-separated
quark and gluon FFs were usually poorly determined for particles
carrying a high fraction of the parton energy. Our measurements in
p+p collisions provide necessary constraints on the FFs in these
ranges, which is crucial for the jet quenching studies at RHIC.
Also shown in Fig.~\ref{ratios} are the $p/\pi^{+}$ and
$\bar{p}$/$\pi^{-}$ ratios in central Au+Au collisions with
central values same as in~\cite{starAuAuPID} and updated
uncertainties at high $p_T$. For $p_T\!>\!6$ GeV/$c$, the errors
of $p/\pi^{+}$ and $\bar{p}$/$\pi^{-}$ in~\cite{starAuAuPID} were
dominated by the systematic uncertainty from the $dE/dx$
calibration, while the uncertainties from the kaon contamination
were estimated to be insignificant with $K^{-}/K^{+}=0.94$ and
$K/\pi$ ratio in the range of 0.16 to 0.20. Although our current
measurement of $K/\pi$ ratio does not rule out this range of 0.16
to 0.20, we re-evaluate the uncertainties in kaon contamination
with the new measurements from p+p collisions and update its error
propagation to the $p/\pi^{+}$ and $\bar{p}$/$\pi^{-}$ ratios in
central Au+Au collisions, shown in Fig.~\ref{ratios}.

\begin{figure}
\includegraphics*[keepaspectratio,scale=0.95]{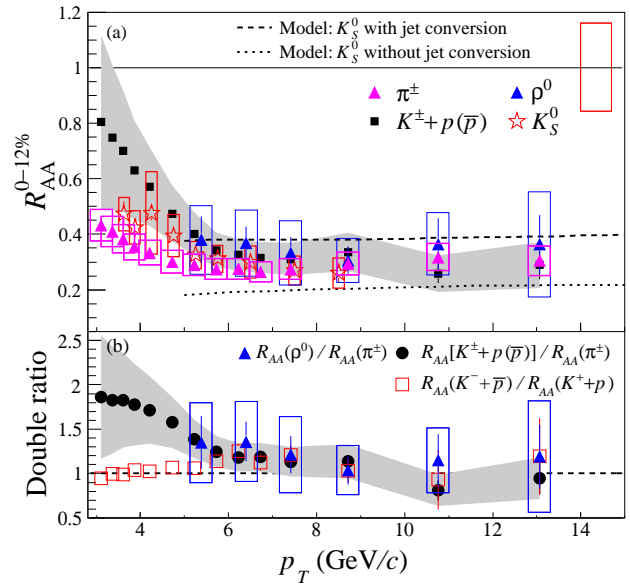}
\caption{(Color Online) (a) $R_{AA}$ of $K^{\pm}+p(\bar{p})$,
$K_S^{0}$, $\rho^{0}$, and $\pi^{\pm}$ in central Au+Au collisions
as a function of $p_T$. The curves are the calculations for
$K_{S}^{0}$ $R_{AA}$ with and without jet conversion in
medium~\cite{weiliu:07}. Bars and boxes (bands) represent
statistical and systematic uncertainties, respectively. The height
of the band at unity represents the normalization uncertainty. (b)
The ratios of $R_{AA}[K^{\pm}+p(\bar{p}),\rho^{0}]$ to
$R_{AA}(\pi^{\pm})$ and $R_{AA}(K^{-}+\bar{p})$ to
$R_{AA}(K^{+}+p)$. The boxes and shaded bands represent the
systematic uncertainties for $R_{AA}(\rho^{0})/R_{AA}(\pi^{\pm})$
and $R_{AA}[K^{\pm}+p(\bar{p})]$/$R_{AA}(\pi^{\pm}$),
respectively. The systematic uncertainties for
$R_{AA}(K^{-}+\bar{p})/R_{AA}(K^{+}+p)$ are 2--12\% and left off
for clarity.} \label{Raa}
\end{figure}

The nuclear modification factors $R_{AA}$ and double ratios of
$R_{AA}$ are shown in Fig.~\ref{Raa} for $K^{\pm}+p(\bar{p})$,
$K_S^{0}$, $\rho^{0}$, and $\pi^{\pm}$. Instead of using the
individually extracted $K$ and $p(\bar{p})$
yields~\cite{starAuAuPID} in the $R_{AA}$, we obtain the combined
$K^{\pm}+p(\bar{p})$ yield with smaller systematic uncertainties
by subtracting the charged pion yields from the inclusive hadron
yields. At $p_{T}{ }^{>}_{\sim}8$ GeV/$c$, a common suppression
pattern is observed for the different mesons ($K_{S}^{0}$,
$\pi^{\pm}$, and $\rho^{0}$), despite the differences in quark
flavor composition and mass. We also observe that $K^{-}+\bar{p}$
shows a magnitude of suppression similar to that of $K^{+}+p$,
despite the different contributions from gluon and quark jets and
any Casimir factor effects on jet energy loss. A model for jet
conversion in the hot and dense medium overpredicts the
$K_{S}^{0}$ enhancement at high $p_T$~\cite{Fries:08}, as shown in
Fig.~\ref{Raa}.

It is worthwhile to highlight two important inputs to the jet
conversion model calculation shown in Fig.~\ref{Raa}: the kaon
spectrum in p+p collisions with the specific FF used in the model
does not match our measurement, and the original
$R_{AA}(K_{S}^{0})$ in the absence of jet conversion was assumed
to be equal to $R_{AA}(\pi^{\pm})$~\cite{Fries:08}.

Enhanced parton splitting can also significantly change the jet
hadron chemical composition~\cite{urs:07}. In this model, heavier
hadrons at high $p_T$ become more abundant relative to the case
without the enhanced parton splitting mechanism. Naively, the
heavier $\rho^{0}$ meson is expected to be less suppressed than
the $\pi^{\pm,0}$ and $\eta$~\cite{phenixeta} since all of them
originate from the same parton fragmentation with similar
constituent quark content. However, our measurements indicate that
the $\rho^{0}$ and $\pi^{\pm}$ suppressions are similar in central
Au+Au collisions. In addition, possible in-medium hadronization in
the deconfined matter can lead to less suppression for protons
than for kaons and pions at $8\!<p_T\!<\!20$
GeV/$c$~\cite{Bellwied:10}. A comprehensive comparison requires
quantitative modeling and calculations incorporating 3D
hydrodynamics in an expanding
medium~\cite{renk:07,arXiv:0906.4231} and proper light
flavor-separated quark and gluon FFs. Since the protons are only a
small part of the inclusive charged hadrons in p+p collisions, we
note that a factor of 2 enhancement of $R_{AA}(p+\bar{p})$
relative to $R_{AA}(\pi^{\pm})$ leads to a 20\% enhancement of
$R_{AA}[K^{\pm}+p(\bar{p})]$ compared to $R_{AA}(\pi^{\pm})$. This
20\% enhancement falls within the range of our systematic
uncertainties~\cite{starppdAuPID}. Improved identified-particle
measurements in Au+Au collisions are needed to tighten constraints
on phenomenological models related to jet quenching.

In summary, we report identified particle $p_T$ spectra at
mid-rapidity up to 15 GeV/$c$ from p+p and Au+Au collisions at
$\sqrt{s_{_{NN}}}$ = 200 GeV. The NLO pQCD models describe the
$\pi^{\pm}$ spectra but fail to reproduce the $K$ and $p(\bar{p})$
spectra at high $p_T$. The measured anti-particle to particle
ratios are observed to decrease with increasing $p_{T}$. This
reflects differences in scattering contributions to the production
of particles and anti-particles at RHIC. At $p_{T}{ }^{>}_{\sim}8$
GeV/$c$, a common suppression pattern is observed for different
particle species. Incorporating our p+p data in generating the
flavor separated FFs in the same kinematic range will provide new
inputs and insights into the mechanisms of jet quenching in heavy
ion collisions.

We thank the RHIC Operations Group and RCF at BNL, the NERSC
Center at LBNL and the Open Science Grid consortium for providing
resources and support. This work was supported in part by the
Offices of NP and HEP within the U.S. DOE Office of Science, the
U.S. NSF, the Sloan Foundation, the DFG cluster of excellence
`Origin and Structure of the Universe' of Germany, STFC of the
United Kingdom, CNRS/IN2P3, FAPESP CNPq of Brazil, Ministry of Ed.
and Sci. of the Russian Federation, NNSFC, CAS, MoST, and MoE of
China, GA and MSMT of the Czech Republic, FOM and NWO of the
Netherlands, DAE, DST, and CSIR of India, Polish Ministry of Sci.
and Higher Ed., Korea Research Foundation, Ministry of Sci., Ed.
and Sports of the Rep. Of Croatia, and RosAtom of Russia.

\end{document}